\documentclass[journal]{IEEEtran}
\IEEEoverridecommandlockouts 
\ifCLASSINFOpdf
\usepackage[pdftex]{graphicx}
\graphicspath{{./Images/}}
\DeclareGraphicsExtensions{.pdf,.jpeg,.png}
\else
\fi

\usepackage[cmex10]{amsmath}
\usepackage{siunitx}
\usepackage{cite}
\usepackage{psfrag}
\usepackage[utf8]{inputenc}
\usepackage[T1]{fontenc}
\usepackage{amsmath,amsfonts,amsbsy,amssymb}
\usepackage{mathabx}
\usepackage{mathrsfs}
\usepackage[nolist]{acronym}
\usepackage{tabularx}
\usepackage{amssymb}
\usepackage{amsmath}
\usepackage{graphicx}
\usepackage{cite}
\usepackage{multirow}
\usepackage{wasysym}
\usepackage{multirow}
\usepackage{float}
\usepackage{xcolor}
\usepackage[font=footnotesize]{caption}
\usepackage{algorithm}
\usepackage{algorithmic}
\usepackage{footnote}
\usepackage{array}
\usepackage[shortcuts,acronym]{glossaries}
\usepackage{xcolor}

\usepackage{glossary-superragged}
\newglossarystyle{modsuper}{%
  \glossarystyle{super}%
  
}

\makeglossaries 

\newacronym{res}{RES}{Renewable Energy Sources}
\newacronym{der}{DER}{Distributed Energy Resources}
\newacronym{mpmc}{MPMC}{Microgrid Protection Management Controller}
\newacronym{ieds}{IEDs}{Intelligent Electronic Devices}
\newacronym{dg}{DG}{Distributed Generation}
\newacronym{cb}{CB}{Circuit Breaker}
\newacronym{ns}{NS}{Network Slicing}
\newacronym{dpn3}{DPN3}{Distributed network protocol}
\newacronym{sntp}{SNTP}{Simple Network Time Protocol}
\newacronym{scada}{SCADA}{Supervisory Control and Data Acquisition}
\newacronym{goose}{GOOSE}{Generic Object-Oriented Substation Event}
\newacronym{rtps}{RTPS}{Real-Time Publish-Subscribe}
\newacronym{dos}{DoS}{Denial of service}
\newacronym{smv}{SMV}{Sampled Measured Values}
\newacronym{lan}{LAN}{Local Area Network}
\newacronym{iot}{IoT}{Internet of Things}
\newacronym{urllc}{URLLC}{Ultra Reliable and Low Latency Communications}
\newacronym{ptp}{PTP}{Precision Time Protocol}
\newacronym{pps}{PPS}{Pulses per Second}

\begin{document}
\title{Review of the State-of-the-Art on Adaptive Protection for Microgrids based on Communications}

\author{Daniel Gutierrez-Rojas, \textit{Student Member, IEEE},  Pedro H. J. Nardelli, \textit{Senior Member, IEEE}, \\Goncalo Mendes, Petar Popovski, \textit{Fellow, IEEE} \thanks{D. Gutierrez-Rojas, P. Nardelli and G. Mendez are with Lappeenranta-Lahti University of Technology, Lappeenranta, Finland. (e-mails: daniel.gutierrez.rojas@lut.fi,Pedro.Nardelli@lut.fi,Goncalo.Mendes@lut.fi). \newline P. Popovski is with the Department of Electronic Systems, Aalborg University, Denmark (e-mail:petarp@es.aau.dk) .}}

\maketitle




\begin{abstract}
The dominance of distributed energy resources in microgrids and the associated weather dependency require flexible protection.
They include devices capable of adapting their protective settings as a reaction to (potential) changes in system state. %
Communication technologies have a key role in this system since the reactions of the adaptive devices shall be coordinated. 
This coordination imposes strict requirements: communications must be available and ultra-reliable with bounded latency in the order of milliseconds.
This paper reviews the state-of-the-art in the field and provides a thorough analysis of the main related communication technologies and optimization techniques.
We also present our perspective on the future of communication deployments in microgrids, indicating the viability of 5G wireless systems and multi-connectivity to enable adaptive protection.
\end{abstract}

\begin{IEEEkeywords}
Microgrids, Adaptive protection, Communication Systems, RES DER, 5G, URLLC. 
\end{IEEEkeywords}


\printacronyms[style=modsuper, type=acronym,nonumberlist, title={List of Acronyms}]

\section{Introduction}

The electrification of energy systems based on \ac{res} contributes towards reaching United Nations Sustainable Development Goal 7 --- \textit{"Ensure access to affordable, reliable, sustainable and modern energy for all"}.
Furthermore, to build transmission lines and distribution lines, as well as new communications infrastructure to serve the traditional power systems, is becoming more and more challenging due to, for instance, growing pressures over environmental licensing, funding allocation, etc.
It has been suggested that the centralized paradigm of energy delivery is reaching its technical boundaries and no longer seems to constitute the most effective approach for granting continuous and reliable power supply to customers located at the edge of the grid, especially in countries with a high percentage of non-urban area installations \cite{Lee2019}.
The above trends have led to increasing interest in installing small scale generation closer to the consumption nodes -- \ac{der}.
%
%
%


Practical modernization of the electrical grid usually refers to small-scale cluster integration of DER and customer demand at the distribution level --- microgrids. Microgrids are localized electrical systems with autonomous control and enhanced grid-demand interaction, which are also able to operate in grid-connected and islanded mode \cite{Marnay2012ServingEA,nardelli2014models}. 
Sophisticated features of microgrids as advanced power electronics and complex control configurations impose substantial technical challenges. Protection schemes and strategies against internal and external faults, which can harm system elements or consumer equipment are among those challenges.
%
Microgrid operational conditions may vary rapidly due to DER contribution with low inertia of non-rotating elements and rapid changes in weather conditions (wind and solar radiation) \cite{Haj-ahmed2013} or due to sudden state changes between connected and islanded mode. 
External faults are normally cleared using conventional protection schemes at the distribution level, but these schemes may not be suitable to microgrid internal faults \cite{Soshinskaya2014}.

To ensure safe and appropriate operation, all variables of the microgrid elements shall be monitored and required changes shall be applied to the device protection settings dynamically when the operating conditions of the grid change (e.g., due to fault occurrence). 
Conventional protection schemes, however, rely on large inertia and long transient periods, which are insufficient in this new microgrid context dominated by DER. 
Thus, adaptive schemes become necessary \cite{Ustun2011,Habib2018}.
The self-implemented changes by adaptive protection devices are based on ``intelligent'' algorithms that process the available data, making the microgrid a cyber-physical system.
This leads to an additional concern about the cyber domain: failures in algorithms may stress or even harm physical components \cite{Cintuglu2017}. 

In microgrids that rely on a central management controller, the communication of \ac{ieds}  is used to keep the system updated on the current state of the grid, tracking the operating currents and making proper fault detection \cite{Habib2018,Wan2010,Habib2017}. 
A reliable communication between the system elements is therefore needed. 
In fact, any type of electrical protection scheme that relies on communication requires robustness, a virtually full-time availability and strictly bounded latency \cite{Baranwal2019}. 
Those stringent requirements associated with communications are hard to meet for any current communication system (either wired or wireless).
{Latency as low as 10 ms}, high reliability (i.e., packet error rate lower than  $99.999\%$), high availability ($\approx$99.999\%) and time synchronization are some of the key requirements that the fifth generation of wireless mobile networks (5G) promise to achieve for safe operation of electrical protection systems and that previous technologies alone cannot satisfy due to lack of performance and cost-effective solutions.
{In particular, the integration of different existing technologies with 5G with other wireless interfaces (e.g., WiFi, LTE, or NB-IoT) to exploit the \textit{interface diversity} also known as multi-connectivity offers an already feasible solution for many applications that requires high reliability with latency at order of milliseconds, as shown in \cite{Nielsen2018}.} 
Such a performance is only becoming possible due to major advancements in machine-type communications, adopting specific solutions for different regimes related to data rates, coverage, availability, reliability and latency.
The deployment of \ac{ns} and different types of control messages to establish connections are also examples of wireless communication engineering solutions to comply with the above mentioned strict  quality of service requirements. 

It is also important to consider the different protocols available for communications in grid protection.
The Standard IEC 61850 includes messaging protocols for control and grid automation that are ideal for adaptive protection.
Although various review papers on adaptive microgrid protection and their communication schemes have been published \cite{Habib2018,Beheshtaein2019a,Beheshtaein2019,Ustun2011}, none of them actually considers the possibility of using emerging 5G mobile communications as part of their proposed solutions. 
We try here to fill this gap by reviewing of the state-of-the-art of adaptive protection focusing on the communication aspects and how 5G technologies can be deployed as an enabling technology.

The rest of this paper is divided as follows.
Section \ref{sec:study_case} presents a generic case that highlights the need for adaptive protection schemes in microgrids. Section \ref{sec:current_adaptive} presents a review of techniques for adaptive protection and communication approaches in microgrids. Section \ref{sec:disc} discusses finding done in previous chapters, introduces how 5G can become a reliable communication system for adaptive microgrid protection and elaborates on outstanding issues and challenges in this area. Conclusions are finally presented in Section \ref{sec:conclusions}.

\section{Adaptive protection schemes in microgrids}
\label{sec:study_case}

The most common type of protection in electrical distribution systems today is overcurrent-based protection.
{This mission-critical application requires from the communication system a latency between 12 and 20 ms with 99.999\% of reliability for sensing/metering and control purposes \cite{Yan2013}. }
Overcurrent protection is impacted more than any other protection function by connection of DER\cite{Moxley2018} due to bidirectional current flow to faulted point. The state of the different circuit breakers in the electrical grid also plays a significant role in the protection settings. Consider a generic case representation of a microgrid depicted in Fig. \ref{fig:study_case} with a common IEC 61850 communication setup.

\vspace{-2ex}
\subsection{Adaptive setting} 

\begin{figure}
    \centering
   \includegraphics[width=\columnwidth]{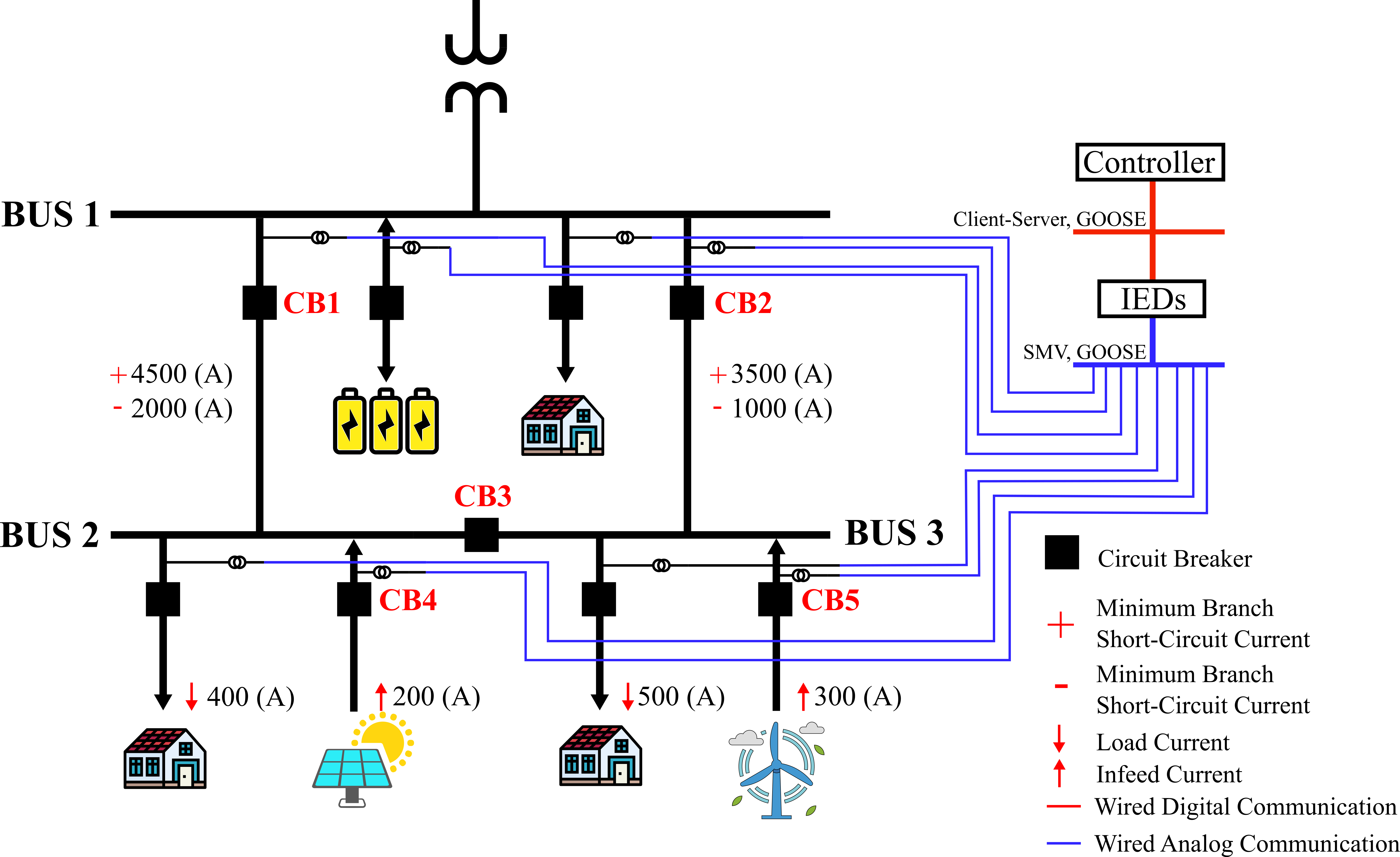}
    \caption{Generic case of a microgrid adaptive setting with fault and load current.}
    \label{fig:study_case}
\end{figure}

The electrical system in Fig. \ref{fig:study_case} is composed by three main circuit breakers (CB1, CB2 and CB3) which are responsible for maintaining the power supply within the microgrid and two circuits breakers (CB4 and CB5) at the DER infeed.
Given overcurrent protection functions for CB1 and CB2 associated with an IED located at BUS 1 and three different cases for their setting and reclosing:

\subsubsection{Case 1: CB1, CB2 Closed and CB3, CB4, CB5 Opened}

Without any infeed from DER at CB4 and CB5, and applying the  rule of thumb where the overcurrent settings ($CB_S$) is inside the interval of double the magnitude of load current $I_l$ and half of the minimum current fault $I_f$, as shown in  \eqref{eq:interval}:

\begin{equation}
CB_S = \left[ I_l\times 2 ,\frac{I_f}{2} \right],
\label{eq:interval}
\end{equation}
where currents are measured in A.

At CB1 the protection setting in relation to the current is given by:
\begin{equation}
CB_{S1} = \left[ 400\times 2 , \frac{2000}{2} \right] \Rightarrow  [800 , 1000].
\label{eq:interval1}
\end{equation}
For CB1, the rule of thumb applies correctly and then we only have to choose a setting value given inside the limits showed in \eqref{eq:interval1}.

Likewise for CB2: 
\begin{equation}
CB_{S2} = \left[ 500\times 2 , \frac{1000}{2} \right] \Rightarrow  [1000 , 500].
\label{eq:interval2}
\end{equation}
In this case, when we do not have an optimal interval, in order to find a setting we sum the minimum fault current $500$ (A) and load current $1000$ (A) divided by $2$, which returns a setting of $750$ (A). The setting must be above load current and below minimal fault current.

\subsubsection{Case 2: CB2, CB3 Closed and CB1, CB4, CB5 Opened}

With CB3 closed, the setting at CB2 has lower margin from minimum fault current due to the increase of load current.
Having $900$ (A) of load current and $1000$ (A) as minimum fault current, we must find a middle point for setting at $950$ (A). As establish before, a setting below the maximum load current could make the protective device operate under normal operating conditions and a setting above minimal current fault the protective device would not be able to identify and clear any fault under faulty conditions. 
This means an increase in the setting at CB2 while the previous setting is inadequate for this case because at some point the load current may be seen as fault current by the IED causing complete isolation of both loads.

\subsubsection{Case 3: CB2, CB3, CB4, CB5 Closed and CB1,Opened}

With the infeed of DER into the microgrid the protection setting at CB2 can also change. 
A total infeed of $500$ A leaves the maximum load seen from the IED at $400$ A and consequently a bigger margin for setting overcurrent protection function at CB2.

These different cases within a simple microgrid configuration shows the necessity of awareness of the IED to know operation conditions of the network so they can adapt to its actual state by changing their overcurrent settings and guarantee a reliable protection for all elements. This means, complete fault isolation including selectivity. Considering case 3 microgrid state, if there is a fault at BUS 2, both loads (or part of the load, if DER had a manageable way to supply part of the load at BUS 3) would get disconnected by operation of CB2, but with a centralized wireless proposed scheme, as shown in the following chapters, that situation could be avoided and power supply of load at BUS 3 could be ensured, by having a lower overcurrent setting at CB2 and operation of CB3 instead.

\subsubsection{Auto-reclosing}

Once a fault in a given microgrid network is cleared by protective devices, it is important to reclose as fast as possible to minimize the lack of power supply and provide stability to the system. 
Auto-reclosing, though, can degrade the life of some elements or even cause permanent damage if the attempt is unsuccessful.
The auto-reclosing action is mostly a control function that can be easily performed at the \ac{mpmc} level, to mitigate any possible damage to the system; the line branches that have less current contribution are the ones to reclose first. 
This implies that the MPMC has to know the current state of the circuit breakers of the microgrid, along with real-time operation currents and fault currents, so that the line branches that reclose first can be determined. 
Since the current measuring is performed at IEDs, these devices need to communicate with the MPMC.
Similarly to the protective system for fault clearance, wireless communication seems to be a more suitable solution for this task due to its flexibility.


\subsection{Adaptive protection algorithms}
Traditional distribution systems are designed to have radial configuration, in order to supply power from a single power source at a time. 
This means that current will flow only in one direction, i.e. from the source (distribution feeders) to the load (consumer). 
Protection functions for radial configuration usually include non-directional overcurrent relays or IEDs, with fixed settings and no need for communication within protective elements \cite{Bansal2019}.
%
%
\begin{figure}
    \centering
    \includegraphics[width=0.9\columnwidth]{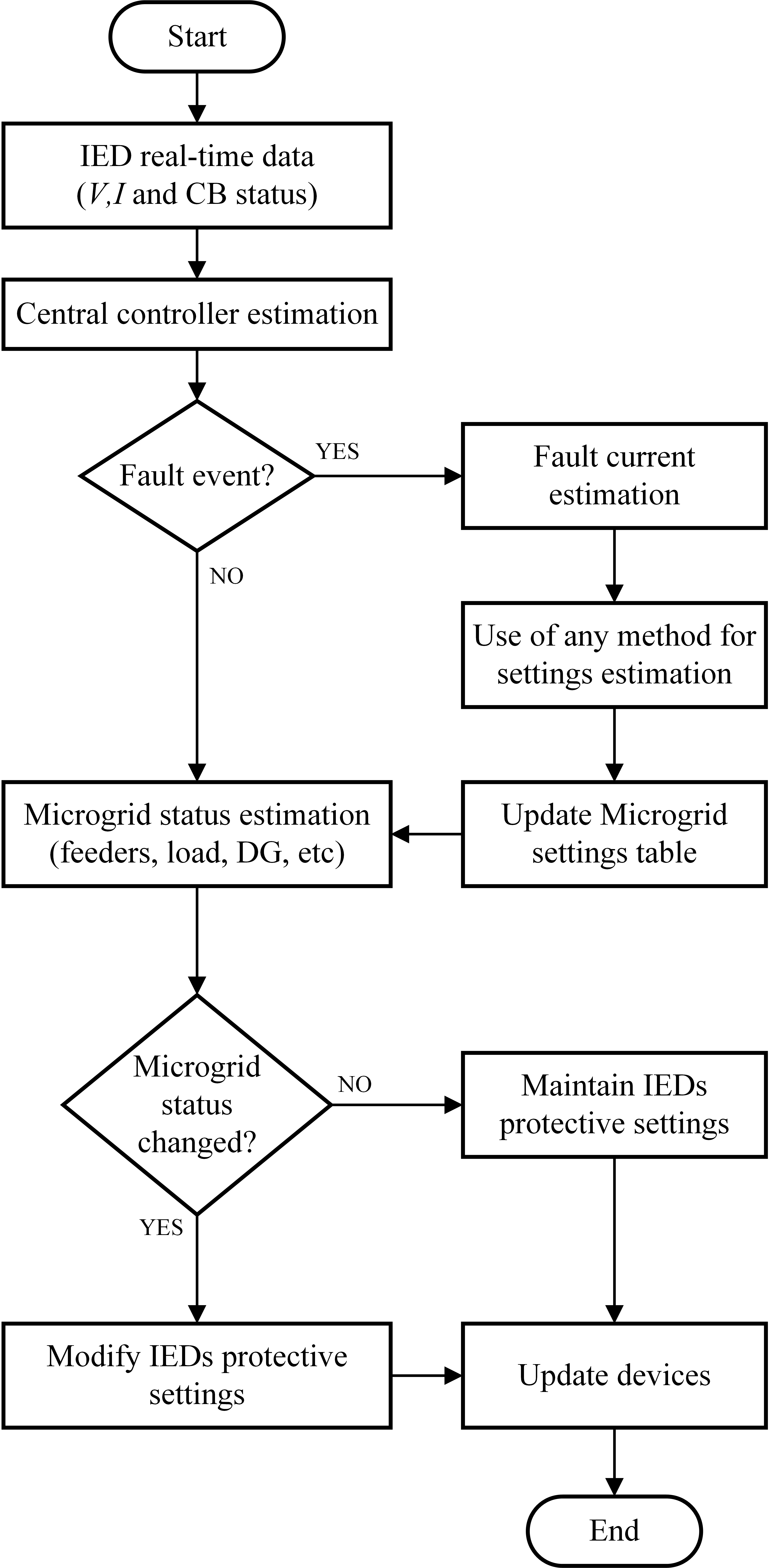}
    \caption{Typical adaptive protection scheme (Adapted from \cite{Ozansoy2015,Lin2019}).}
    \label{fig:flow_chart}
    \vspace{-2ex}
\end{figure}
As microgrids start to proliferate and DER penetration in distribution networks increases, power flow and therefore also fault current become bidirectional. 
Adaptive protection schemes appear as an option to solve the fault clearance challenges that are imposed in this scenario.

\begin{figure}
    \centering
    \includegraphics[width=\columnwidth]{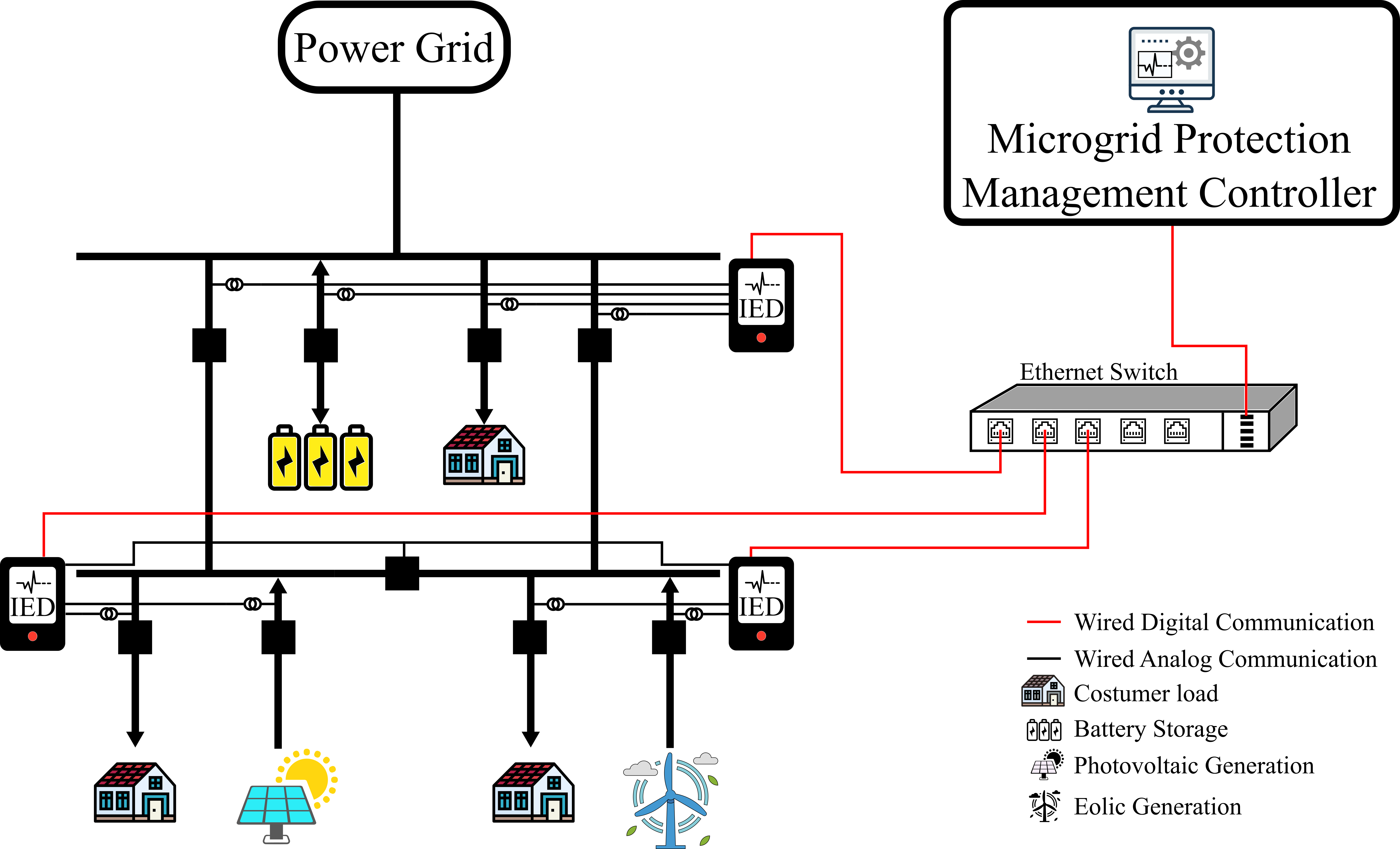}
    \caption{Implementation of wired ethernet-base communications for an overcurrent adaptive protection scheme (Adapted from \cite{Ozansoy2015}).}
    \label{fig:ethernet}
    \vspace{-2ex}
\end{figure}

Fig. \ref{fig:flow_chart} shows the flow chart of a typical adaptive protective scheme implementation. 
First the real-time data gathered by the IEDs is collected and sent through a wired communication channel (usually Ethernet-based) where it is received by the MPMC (Fig. \ref{fig:ethernet}) \cite{Jin2018a}, which will analyse if a trip action was made and whether it was, or not, from a fault occurrence. 
Then, the microgrid state is evaluated for possible temporary conditions in the system after any possible reclose from the circuit breakers. 
Based on the fault currents the system will update the settings at the decision-making table and depending of the state of circuit breakers, a signal could be sent back to the IEDs to rewrite their actual settings for the new ones. 

Additionally, in \cite{Lin2019} after the measurements are gathered, a block of artificial neural networks and another of support vector machine algorithms estimate whether there is a fault and its location, respectively. 
A least square estimation is employed for comparison before updating the decision table.
In \cite{Daryani2018}, the whole tripping process is shown by dividing the flow chart into two main blocks (relay agent and central controller agent) performing an examination of grid state and updating the values of relays. After a fault occurs, the new state is evaluated to calculate new relay settings.
A calculation of the average of total communication latency that involves the previous described blocks was described in \cite{Jin2018a}.
Adaptive protection schemes use different methods to solve their setting adjustment when needed.
Those methods also rely on different optimization techniques to find an efficient but fast method to change a predetermined variable of the IED.
Examples include differential search algorithm \cite{Singh2016}, fuzzy logic and genetic algorithm \cite{Naily2018}, modified particle swarm optimization \cite{Atteya2017}.

\section{Existing communication approaches in adaptive protection systems}
\label{sec:current_adaptive}

\subsection{Wired and wireless implementations}

In wired communication-based automation and adaptive protection implementations, the data transfer between IEDs and the MPMC takes place through cables installed at the substation level. Wireless communication, on the other hand, operates based on radio frequency signals. Both implementations have advantages and disadvantages and whether one is more appropriate than the other is entirely reliant of the use case. Table \ref{ta:advan} presents a comparison between wired and wireless applications of some of the characteristics of substation control that are relevant for adaptive protection. 

\begin{table}
\caption{Wired and wireless communication for substation automation.}
\label{ta:advan}
\begin{tabularx}{\columnwidth}{p{1cm}|X|X}

\multicolumn{1}{c|}{\textbf{Characteristics}} & \multicolumn{1}{c|}{\textbf{Wired}}                                                                  & \multicolumn{1}{c}{\textbf{Wireless}}                                                              \\ \hline\hline
Reliability                                    & - Once the installation is complete, probability to fail is very low                                 & - Redundancy can lower probability to fail                                                         \\ \hline
Stability                                      & - Not distorted by other connections or objects                                                      & - Variation in the latency could be experienced depending on the interference by other networks    \\ \hline
Visibility                                     & - Not visible by other wired connections but could be connected by nodes to facilitate data transfer & - Might be visible to other wireless connections depending of the bandwidth                        \\ \hline
Speed                                          & - Independent cables avoid unexpected and unnecessary data making transfer faster                    & - Latency of 5G deployments can perform equal or better than wired networks             \\ \hline
Security                                       & - Firewall and other applications provide enough security when the installation is monitored         & - Signals that propagate through can be intercepted. Proper encryption technologies can avoid this \\ \hline
Cost                                           & - Design, space adequation and installation could be costly                                          & - Cost of installation relatively inexpensive                                                      \\ \hline
Mobility                                       & - Stationary without possibility of fast reallocation                                                & - Flexible and easy to add new components or reallocation                                          \\ \hline
Installation                                   & - Depending on size and requirements, it can take longer to set up                                  & - Requires less equipment and fast installation                                                    \\ \hline
Maintenance                                    & - Potentially costly depending of number of elements                                                 & - Due to less elements, less costly and less frequent maintenance                                 
\end{tabularx}
\end{table}

Wired connections are generally considered to be highly reliable but their total cost and lack of flexibility impose additional challenges when new equipment is installed at the substation. Wired and wireless communication can also be combined to enhance the tasks performed by each element of the network, such as in \cite{Sotomayor2018}, where a mix of technologies such as Fiber Optics, Broadband Power Line over medium voltage, and Wi-Fi are used for control and measuring. However, most work  found in the literature adopts less sophisticated physical wired communications, for high reliability and low latency.

In this context, the role of emerging technologies in wireless communications (5G and integration of 5G other wireless communication interfaces) can be groundbreaking. Not only will these be able to efficiently address the drawbacks from legacy wireless communications, but also to significantly enhance its capabilities. Furthermore, the discussion on the need for more versatile communication technologies i.e. applicable to the generality of implementation use cases, increasing efficiency and reducing costs, is a valid one. Thus, the authors propose a change of paradigm of microgrid automation and control towards a scenario of prevalent adaptive protection implementations, which as explained constitute a significant departure from contemporary wired installations.


\begin{table*}[!htbp]
\begin{center}
\caption{Mapping of communication approaches used in adaptive protection schemes for microgrids. Their main features are discussed throughout Sec. \ref{sec:current_adaptive}.}
\label{ta:adaptive}
\begin{tabular}{|c|c|c|c|c|c|c|c|c|}
\hline
\multicolumn{2}{|c|}{Reference}                                                                             & \multicolumn{2}{c|}{Controller} & \multicolumn{3}{c|}{Communication}              & \multicolumn{2}{c|}{Operation Mode} \\ \hline
Year                   & Cite                                                                               & Centralized   & Decentralized   & Wired      & Wireless   & Standard/Protocol     & Grid-connected     & Islanded       \\ \hline
\multirow{8}{*}{2019}  & \cite{Lin2019}                                                                     & \checkmark    &                 & \checkmark &            & IEC 61850, SNTP                  & \checkmark         & \checkmark     \\
                       & \cite{Alam2019a}                                                                   & \checkmark    &                 & \checkmark &            & IEC 61850             & \checkmark         &                \\
                       & \cite{Teimourzadeh2019}                                                            & \checkmark    &                 & ---        & ---        & ---                   & \checkmark         &                \\
                       & \cite{Singh2019}                                                                   &               & \checkmark      & \checkmark &            & --- & \checkmark         & \checkmark     \\
                       & \cite{Nougain2019}                                                                 & \checkmark    &                 & \checkmark &            & ---                   & \checkmark         &                \\
                       & \cite{Habib2019}                                                                   & \checkmark    &                 & \checkmark &            & RTPS                  &                    & \checkmark     \\
                       & \cite{Momesso2019}                                                                 &               & \checkmark      &            &            &                       & \checkmark         &                \\
                       & \cite{Ferreira2019}                                                                &               & \checkmark      &            &            &                       & \checkmark         & \checkmark     \\ \hline
\multirow{16}{*}{2018} & \cite{Hosseini2018, Paladhi2018a}                                                  &               & \checkmark      & ---        & ---        & ---                   & \checkmark         &                \\
                       & \cite{AsghariGovar2018}                                                            &               & \checkmark      & \checkmark &            & ---                   & \checkmark         &                \\
                       & \cite{Chandraratne2018,Sedghisigarchi2018,Amaratunge2018}                          & \checkmark    &                 & ---        & ---        & IEC 61850, DPN3 & \checkmark         & \checkmark     \\
                       & \cite{Shah2018}                                                                    & \checkmark    &                 & \checkmark &            & IEC 61850             & \checkmark         & \checkmark     \\
                       & \cite{Peiris2018}                                                                  & \checkmark    &                 & \checkmark &            & ---                   & \checkmark         & \checkmark     \\
                       & \cite{daSilva2018}                                                                 & \checkmark    &                 & \checkmark &            & Telnet                & \checkmark         &                \\
                       & \cite{GhaleiMonfaredZanjani2018}                                                   & \checkmark    &                 & \checkmark &            & ---                   & \checkmark         &                \\
                       & \cite{Jin2018a}                                                                &               & \checkmark      & \checkmark &            & IEC 61850             & \checkmark         &                \\
                       & \cite{Xu2018,Xu2018b}                                                              & \checkmark    &                 &            & \checkmark & IEC 61850,60870-5-101 & \checkmark         &                \\
                       & \cite{Daryani2018}                                                                 & \checkmark    & \checkmark      & \checkmark &            & IEC 61850             & \checkmark         & \checkmark     \\
                       & \cite{Naily2018}                                                                   & \checkmark    &                 & ---        & ---        & IEC 61850,60870-5-101 & \checkmark         &                \\
                       & \cite{Habib2018}                                                                   &               & \checkmark      & \checkmark &            &                       &                    & \checkmark     \\
                       & \cite{Sotomayor2018}                                                               &               & \checkmark      & \checkmark & \checkmark & IEC 61850             & \checkmark         &                \\
                       & \cite{Zhonghua2018}                                                                &               & \checkmark      & ---        & ---        & ---                   & \checkmark         & \checkmark     \\
                       & \cite{Ma2018}                                                                      & ---           & ---             & ---        & ---        & ---                   & \checkmark         &                \\
                       & \cite{CAI2018,Kang2018,Linli2018,Upadhiya2018,He2018,Elhadad2018,Zhao2018}         &    ---   &        ---      & ---        &     ---       &     ---                  &      ---              &      ---          \\ \hline
\multirow{10}{*}{2017} & \cite{Habib2017a}                                                                  & \checkmark    & \checkmark      & ---        & ---        & ---                   & \checkmark         & \checkmark     \\
                       & \cite{Purwar2017}                                                                  & \checkmark    &                 & ---        & ---        & ---                   & \checkmark         & \checkmark     \\
                       & \cite{Pacheco2017}                                                                 & \checkmark    &                 & ---        & ---        & IEC 61850, DPN3       & \checkmark         & \checkmark     \\
                       & \cite{Wheeler2017}                                                                 &               & \checkmark      & \checkmark &            & ---                   & \checkmark         &                \\
                       & \cite{Naily2017}                                                                 &               & \checkmark      &  &  \checkmark          & ---                   & \checkmark         &                \\
                       & \cite{Gaber2017}                                                                   & \checkmark    & \checkmark      & ---        & ---        & ---                   & \checkmark         & \checkmark     \\
                       & \cite{Orji2017}                                                                    &               & \checkmark      & \checkmark &            & ---                   & \checkmark         & \checkmark     \\
                       & \cite{Piesciorovsky2017}                                                           & \checkmark    &                 & \checkmark &            & ---                   & \checkmark         &                \\
                       & \cite{Tang2017}                                                                    &               & \checkmark      & \checkmark &            & Point-to-Point                     & \checkmark         &                \\
                       & \cite{Muda2017}                                                                    & \checkmark    &                 & \checkmark &            & ---                   & \checkmark         & \checkmark     \\
                       & \cite{Piesciorovsky2017a,Shen2017,George2017,Atteya2017,Song2017,Tjahjono2017}     & ---           & ---             & ---        & ---        & ---                   & \checkmark         &                \\ \hline
\multirow{10}{*}{2016} & \cite{Misak2016,Sanca2016,Leite2016}                                               & \checkmark    &                 & \checkmark &            & ---                   & \checkmark         &                \\
                       & \cite{Liu2016}                                                                     & \checkmark    &                 & \checkmark &            & IEC 61850             & \checkmark         & \checkmark     \\
                       & \cite{HengweiLin2016}                                                              & \checkmark    &                 & \checkmark &            & IEC 61850, DPN3  & \checkmark         & \checkmark     \\
                       & \cite{Singh2016}                                                                   & \checkmark    &                 & \checkmark &            & IEC 61850             & \checkmark         &                \\
                       & \cite{Shih2016}                                                                    & \checkmark    &                 & \checkmark &            & IEC 61850, DPN3       & \checkmark         &                \\
                       & \cite{Bari2016}                                                                    & ---           & ---             &            & \checkmark & ---                   & \checkmark         &                \\
                       & \cite{GnanaSwathika2016}                                                           & \checkmark    &                 & ---        & ---        & IEC 61850, IEEE 1588  & \checkmark         & \checkmark     \\
                       & \cite{Muda2016}                                                                    & \checkmark    &                 & ---        & ---        & ---                   & \checkmark         & \checkmark     \\
                       & \cite{Pujiantara2016}                                                              & ---           & ---             & ---        & ---        & ---                   & \checkmark         & \checkmark     \\
                       & \cite{Shichen2016,He2016,Grebchenko2016,Bujanovic2016,XiaobinGuo2016,Jiandong2016} & ---           & ---             & ---        & ---        & ---                   & \checkmark         &                \\ \hline
\multirow{12}{*}{2015} & \cite{DellaGiustina2015}                                                           & \checkmark    &                 & \checkmark &            & IEC 61850             & \checkmark         &                \\
                       & \cite{Lin2015}                                                                     & \checkmark    &                 & ---        & ---        & IEC 61850        & \checkmark         & \checkmark     \\
                       & \cite{Vijitha2015}                                                                 & \checkmark    &                 & ---        & ---        & ---                   & \checkmark         & \checkmark     \\
                       & \cite{Farsadi2015}                                                                 & \checkmark    &                 &            &            &                       & \checkmark         & \checkmark     \\
                       & \cite{Papaspiliotopoulos2015,Gupta2015}                                            & \checkmark    &                 & \checkmark &            & ---                   & \checkmark         &                \\
                       & \cite{Fan2015}                                                                     &               &                 &            &            &                       & \checkmark         &                \\
                       & \cite{Ferreira2015}                                                                &               &                 &            &            &                       &                    & \checkmark     \\
                       & \cite{George2015,Nascimento2015}                                                   & \checkmark    &                 & ---        & ---        & ---                   & \checkmark         &                \\
                       & \cite{Ozansoy2015a,Coffele2015,Ozansoy2015}                                        & \checkmark    &                 & \checkmark &            & IEC61850              & \checkmark         & \checkmark     \\
                       & \cite{Bhattarai2015}                                                               & \checkmark    & \checkmark      & ---        & ---        & ---                   & \checkmark         & \checkmark     \\
                       & \cite{Tummasit2015}                                                                & ---           & ---             & ---        & ---        & ---                   & \checkmark         & \checkmark     \\
                       & \cite{Esmaili2015,Kumar2015,Lopez2015}                                             & ---           & ---             & ---        & ---        & ---                   & \checkmark         &                \\ \hline
\end{tabular} \\
\end{center}
\hspace{1cm}\footnotesize{--- Not specified}
\end{table*}
\begin{figure*}
  \includegraphics[width=\textwidth]{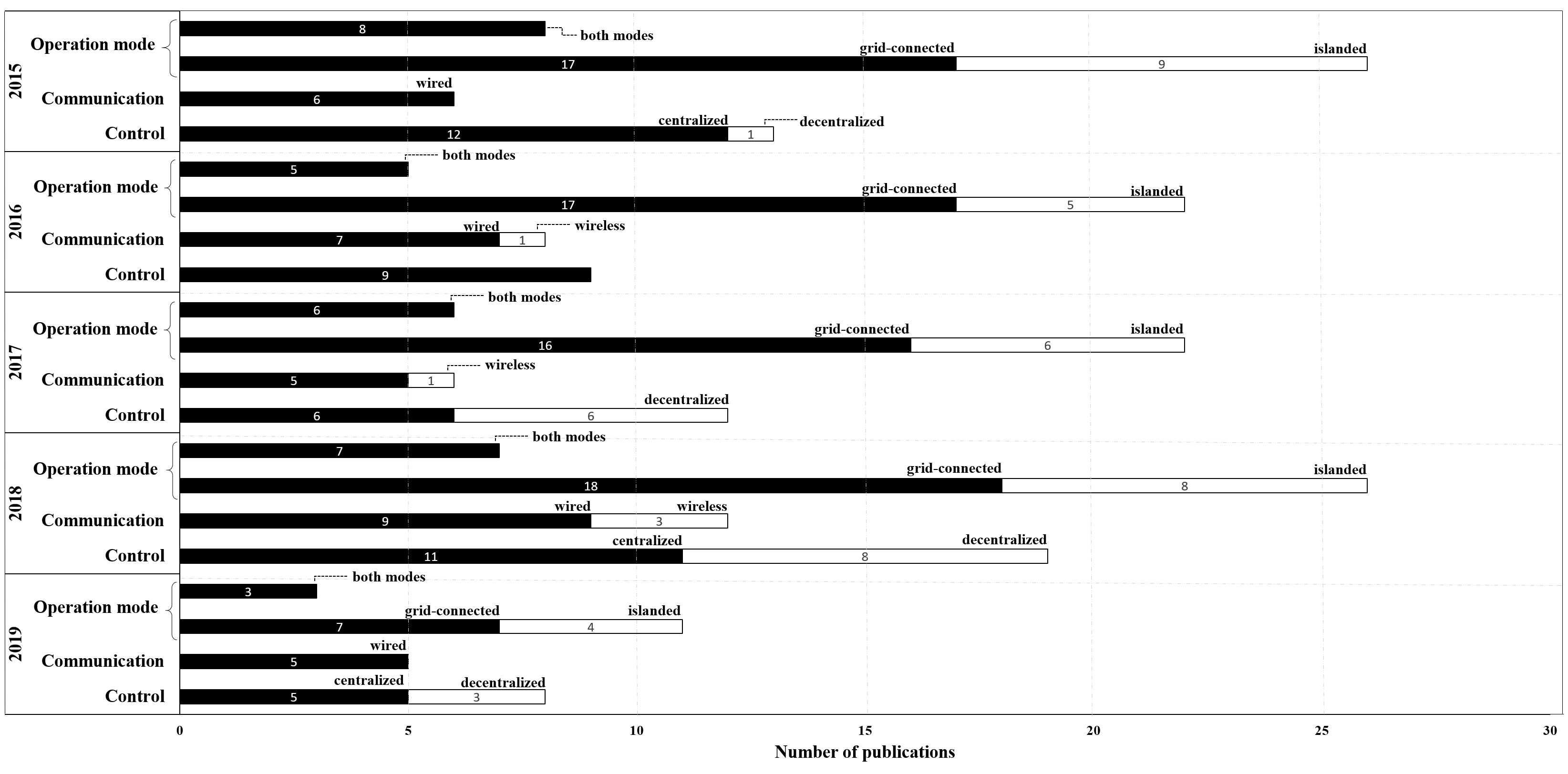}
  \caption{Communication approaches found in microgrid adaptive protection literature, expressed in number of publications per year.}
  \label{fig:bar}
\end{figure*}
\subsection{Traditional communication architectures}
Recent literature on adaptive protection of microgrids has revealed a variety of approaches for analyzing the performance of the respective algorithms and methodologies.
Some approaches focus on centralized or decentralized management for data processing and control, while others focus on the communication infrastructure to reduce times of on-line settings adjustment.
Most of the utilized algorithms were tested in grid-connected operation conditions. A small set, however, can also work under islanded mode, in order to test control robustness of adaptive protection in case of communication failures or disconnection from the grid, when DER are present.

Table \ref{ta:adaptive} summarizes the aforementioned approaches to adaptive protection in microgrids, in the last five years.
In \cite{Lin2019}, a centralized approach chosen.
The paper states that the methodology requires database available before hand and it is obtained through simulation.
It proposes a data mining methodology to quantitatively extract meaningful information from the database. 

As for the implementation, the authors used a wired communication approach, along with \ac{sntp} and \ac{scada}, which includes the IEC 61850 standard.
The authors considered both grid connected and island operation modes.
A fractionalization of microgrid protection is made in \cite{Singh2019} to avoid dependency of centralized management and to improve reliability, which can also work in grid-connected and island operation modes.
In \cite{Jin2018a} and \cite{Sotomayor2018}, a decentralized methodology is proposed using the IEC 61850 standard for grid-connected operation mode.
A combination of adaptive communication-based decentralized (pre-contingency) and centralized (post-contingency) protection schemes is shown in \cite{Daryani2018}, which is suitable for both grid-connected and islanded operation modes. Also in this paper, the IEC 61850 is used for communication between the elements.

When a microgrid is in island mode, it often looses its communication capabilities with a central server, leaving all  protection devices operating with stationary settings or not being adjusted to the lower setting, which means the fault will not be detected. 
To overcome this problem, in case of communication failure, \cite{Habib2017a} proposes a solution using a supercapacitor with bidirectional Voltage Source Converter to contribute for the fault current and raise current value to certain level, which is sensed by the relay and a comparison between high and low settings can be made. 
In \cite{Naily2017}, numerical relays and a global system for Mobile communications modem are connected to communicate with each other (schematic shown in \cite{Elhaffar2015}) and perform a decentralized adaptive protective action due to very good coverage. 
Also in \cite{Xu2018}, the authors propose a SCADA system with Advanced Meter Infrastructure (AMI) and 4G wireless communication. 

The SCADA system is used to perform the online adaptive feature, by obtaining measurements from DER output and AMI.
To acquire the mentioned data from the distribution system to the control center, a 4G wireless communication system was used. 
Lastly, in their work, \cite{Bari2016} suggest that the information exchange between the elements can be accomplished by a Wireless Sensor Network.

Fig. \ref{fig:bar} offers a quantitative analysis of the communication approaches used in adaptive protection of microgrids in recent literature, based on 85 compiled papers  from the last five years. The analysis is expressed in terms of communication technology (wired or wireless), control approach (centralized or decentralized) and operation mode (grid-connected, islanded, or both operation modes). It is important to make the remark that the literature review spans from January 2015 to July 2019 i.e. publications compiled for 2019 do not reflect an entire year. The findings from this analysis are further discussed in Section \ref{sec:disc}.

\subsection{Communication standards and protocols for substation automation and control}
When it comes to communications architecture, the IEC 61850 is a widely accepted standard for automation and equipment of power utilities and DER, specifically for defining protocols for IEDs at electrical substations \cite{IEEEStand}. There are three main protocols defined by the IEC 61850: 
\begin{itemize}
    \item \textbf{\ac{goose}:} Used to send data from IED to IED or from IED to circuit breakers due to its high-speed and high priority characteristics, suitable for tasks such as command trips or alarms;
    \item \textbf{\ac{smv}:} Used to transfer the analog channels of current and voltage to the IED;
    \item \textbf{Manufacturing message specification:} Used for applications that are non-time-critical, such as communications between controller and between substations.
\end{itemize}  

IEC 61850 also defines generic substations events which is a control model that provides a fast and reliable mechanism for data transferring over the electrical substation network. The generic substations events model is divided into earlier described GOOSE and generic substation state events.
All of the above tasks, performed inside communication layers within a power system, are adequate for protection-related applications.
The three protocols run over Transmission Control Protocol, Internet Protocol or a \ac{lan} that can use high speed switched Ethernet like in \cite{Ozansoy2015}. 

IEC 61850 entails additional features, such as data modelling, reporting schemes, fast transfer of events, setting groups, sampled data transfer, commands and data storage, which justify its use in substations and grid protection.
A communication setup using IEC 61850 standard makes it relatively simple to achieve low latency, normally around 4 milliseconds, which is ideal for protection purposes.
Although many of the current implementations using this standard use wired Ethernet or Fiber Optics physical layers, wireless communication may also be implemented using IEC 61850 for communications between the substation elements. 

Other standards used are for instance the IEEE 1588, which describes a hierarchical master-slave architecture for clock distribution and introduces precision time protocol (PTP), used to synchronize clocks throughout a computer network. On a local area network, it achieves clock accuracy in the sub-microsecond range, making it suitable for measurement and control system applications \cite{IEEE1588}. 

PTP supports the transmission of GOOSE messages over an Ethernet network using IEC  61850.  This is generally implemented in SCADA systems where several substations can be covered.  For instance, reference \cite{ptp2009} shows that monitoring three pulses per second (PPS) signals from master to slave  can be synchronized within 200 ns and deliver accurate time stamps below 500 ns.   Note that this delay has a much lower order of magnitude compared to the adaptive protection needs (order of milliseconds), making them negligible.
Also, the IEC 60870-5 defines systems used for telecontrol, supervisory control and data acquisition in electrical engineering and power system automation applications. It provides the communication architecture for sending basic telecontrol messages between two elements (ex. IED and MPMC) that have permanent connected communication channels. IEC 60870-5-101 specifically refers to companion standards for basic telecontrol tasks, which are commonly used in substation control and protection in SCADA systems.
\begin{figure}
  \includegraphics[width=\columnwidth]{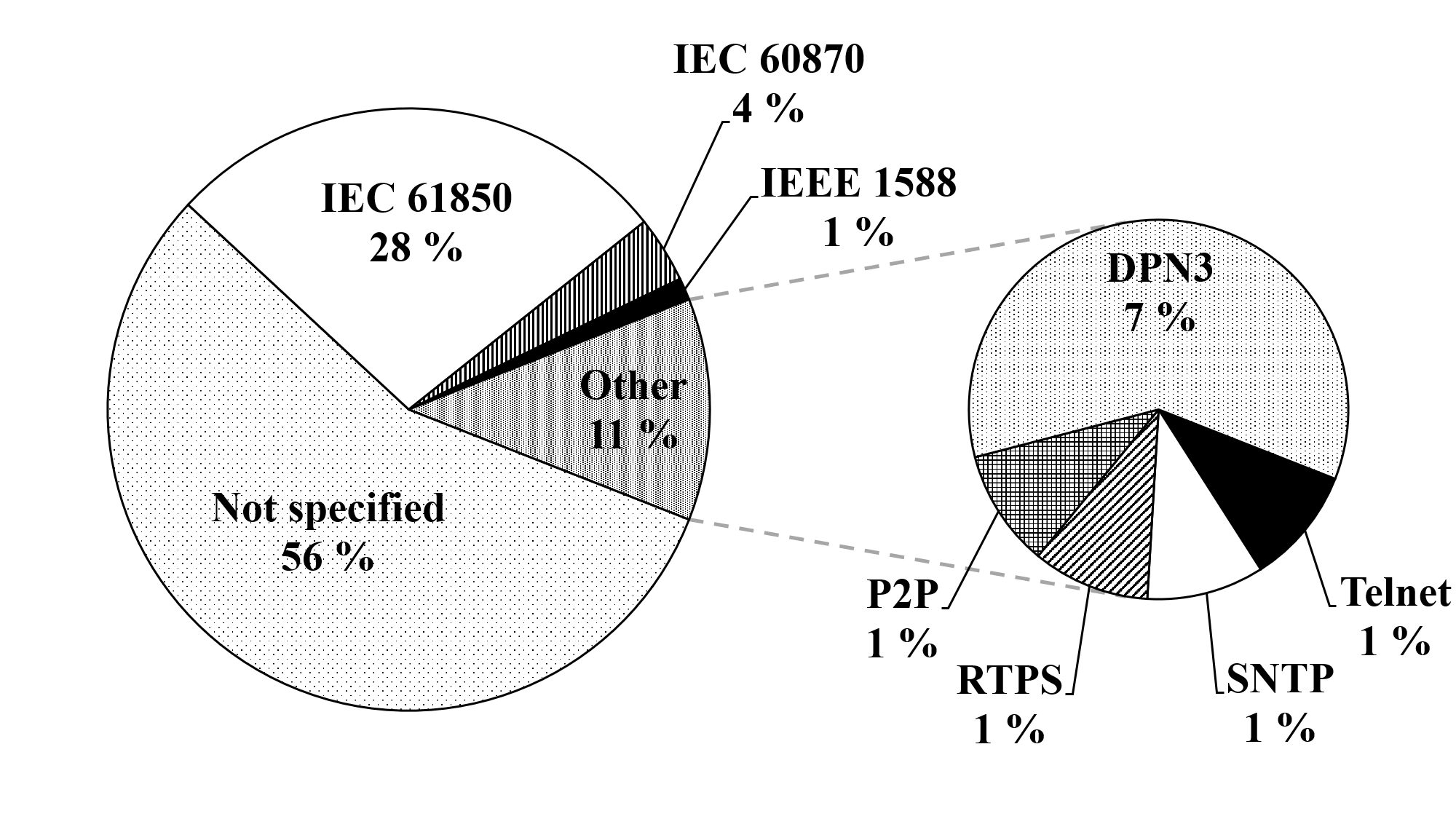}
  \caption{Percent distribution of communication standards and protocols used in microgrid adaptive protection literature.}
  \label{fig:pie}
\end{figure}

Other protocols used for control purposes found in the literature and listed in Table \ref{ta:adaptive} are:
\begin{itemize}
    \item \textbf{\ac{dpn3}}: Used  mainly for communication between a master and remote terminal unit or IEDs.  It provides multiplexing, data fragmentation, error checking, link control, prioritization, and layer 2 addressing services for user data. The protocol is robust, efficient and compatible with many elements which is suitable for SCADA systems. Depending on the elements and the applications it can become very complex;
    \item \textbf{Telnet}: Application protocol used in internet or LAN to provide interactive text-oriented communication systems using a virtual terminal connection and data being interspersed in-band with control information over 8-byte transmission control protocol. Telnet was often used to perform remote connection applications. It doesn't use, however, any form of encrypting mechanism, which makes it vulnerable in modern security terms; 
    \item \textbf{\ac{rtps}}: Protocol which provides two main communication models, the publish-subscribe protocol that transfers data from publishers to subscribers, and the composite State Transfer protocol that transfers states. It features characteristics such as modularity, scalability and extensibility and it's suitable for real time applications running over standard internet protocol networks;
    \item \textbf{Peer-to-peer}: Allows to connect a large number of users over a LAN. The scalability is no longer limited by the server. Its functions are distributed among a number of client peers, communicating in multicast mode. Messages are sent from one client directly to another client, without relying on a central server.
\end{itemize}

Fig. \ref{fig:pie} shows the percent distribution of communication standards and protocols used in recent microgrid adaptive protection literature (based on the same 85-research paper sample). An immediate observation is the dominance of the IEC 61850 standard, which suggests its protocols are suitable for adaptive protection tasks even in the case of wireless deployments, as showed in Table \ref{ta:adaptive}. 

One additional consideration to communication standards and protocols is the physical capability of network elements. 
Adaptive protection requires robust and flexible elements for data gathering and control. Due to their ability to receive and send data to form the closed loop of the adaptive process, IEDs comply fully with these strict requirements.
IEDs must also count with sufficient flash memory capabilities to read/write protective settings \cite{Moxley2018} and successfully achieve the communication data exchange. Reference \cite{Bari2016} also mentions that IEDs should have the ability of logging voluminous information about system parameters.
In \cite{Xu2018,Xu2018b}, the authors selected the most suitable wireless technology for collecting data in real time and transfer it to the central controller, based on synergies with SCADA systems.
\vspace{-2ex}
\subsection{Cyber-security}

The transition of microgrids to the cyber-physical domain comes with a number of cyber-security risks. 
Communication systems are vulnerable to malicious cyber-attacks. If the protection systems in place do not perform appropriately, such attacks can potentially harm the physical domain  \cite{Beheshtaein2019a}. 
Cyber-attacks can be classified in two main categories: Network Security attacks and Goose \& SMV message attacks \cite{Habib2018}.
Three types of attacks related to Network Security are: 
\begin{itemize}
\item \textbf{\ac{dos}}: DoS prevents authorized users to access a service and affects the timeliness of the information exchange, which can cause packet losses. \cite{Peng2017} addresses the case of load frequency control in a power system where  supply is limited from DoS attacks by transferring the model of multi-area power systems to a dependent time delay model, in order to tolerate a certain degree of data losses induced by energy-limited DoS. Many classical approaches address this type of attacks by using distribute topology formation techniques that are based upon the cooperation between IED nodes \cite{dos2015};
\item \textbf{Password cracking attempts}: This method is based on attempts to gain access to system devices (such as IEDs) to gain control over them, performing tripping actions or blocking them from protective signals. For techniques to detect type of attacks, see \cite{Hong2019};
\item \textbf{Eavesdropping attacks}: This type of attack is done by accessing the communication link between the control center and the substation, and can be performed in both wired and wireless communication implementations. The data packets are intercepted by the intruder, who is able to replace real data for fabricated one. After, the controller can send back to the IEDs tripping signals out of wrong information provided by the intruder \cite{Zhang2015}.
\end{itemize}

For GOOSE \& SMV attacks, we have:
\begin{itemize}
\item \textbf{Goose \& SMV modification attacks}: In this type of attack, the intruder modifies the message data between the IED (GOOSE sender) and the circuit breaker (GOOSE receiver) without any notice. And as SMV the intruder can send wrong information about the analog variables of the system. In \cite{Wright2018}, a case where the minimum capabilities an intruder needs to inject a single message and perform undesirable actions is presented;
\item \textbf{Goose \& SMV DoS attacks}: The intruder can prevent the correct operation of the IED by sending a great amount of messages to a IED target causing communication collapse and making it unable to respond to other messages;
\item \textbf{Goose \& SMV replay attacks}: Fault information packets are kept from the intruder and then sent back to the elements under normal operation, causing undesirable tripping and possible substation outages.
\end{itemize} 

When a communication failure resulting from cyber-attacks takes place in a microgrid, it would usually trigger microgrid islanding, which poses challenges to protective devices.
\cite{Habib2018} envisions such a scenario, devising an approach to handle relying on energy storage. 
Under service of energy storage, the IEDs may be able to reach the overcurrent fixed setting to perform tripping actions in case of fault condition, guaranteeing protection actuation and therefore no damages to the microgrid.
 
The literature is abundant in terms of proposed approaches for evaluating and preventing cyber-attack in electrical networks \cite{Rahman2017}. However, for sake of effectiveness and robustness of operations, cyber-security should be approached holistically and from a project design stage. 
Therefore, to prevent those attacks, guaranteeing a reliable cyber-physical protective system embedded in the communication architecture of microgrids, substantial improvements, and thus investments in prevention, detection, mitigation and resilience must still be undertaken.

\vspace{-2ex}
\section{Discussion, open issues and challenges}
\label{sec:disc}
The increasing penetration of RES in electrical networks and the dissemination of microgrids are generating interest in developing communication technologies tailored to new uses and functionalities. For instance, islanded operation will become more relevant (as seen in Fig. \ref{fig:bar}), driving the need for further adaptability in protective units for system elements. 
Unprecedented changes have taken place in the ways in which people communicate during the last two decades.
Changes in the communication infrastructure of distribution systems and microgrids are also important and ruled by the need for greater flexibility and more cost-effective solutions. The research presented in this paper highlights the predominance of wired, centralized communication approaches for adaptive protection in microgrids. On the other hand, it reveals no identifiable changing trend in terms of adopted communication technology (wired or wireless) in recent practical and theoretical research (Fig. \ref{fig:bar}).
There is a dominant use of IEC 61850 standard because it addresses necessary communication protocols in the substation domain \ref{fig:pie}. IEC 61850 is suitable for wireless communications and can be used for future implementation of protection and control systems.
Many further developments such as the \ac{iot}, augmented reality, telemedicine, virtual reality and unmanned driving, have been applied to real businesses.
These developments have brought significant changes to society and their mobile communication requirements became higher \cite{Nardelli2019,Wang2017,Durisi2016}. 

\begin{figure}
    \centering
    \includegraphics[width=\columnwidth]{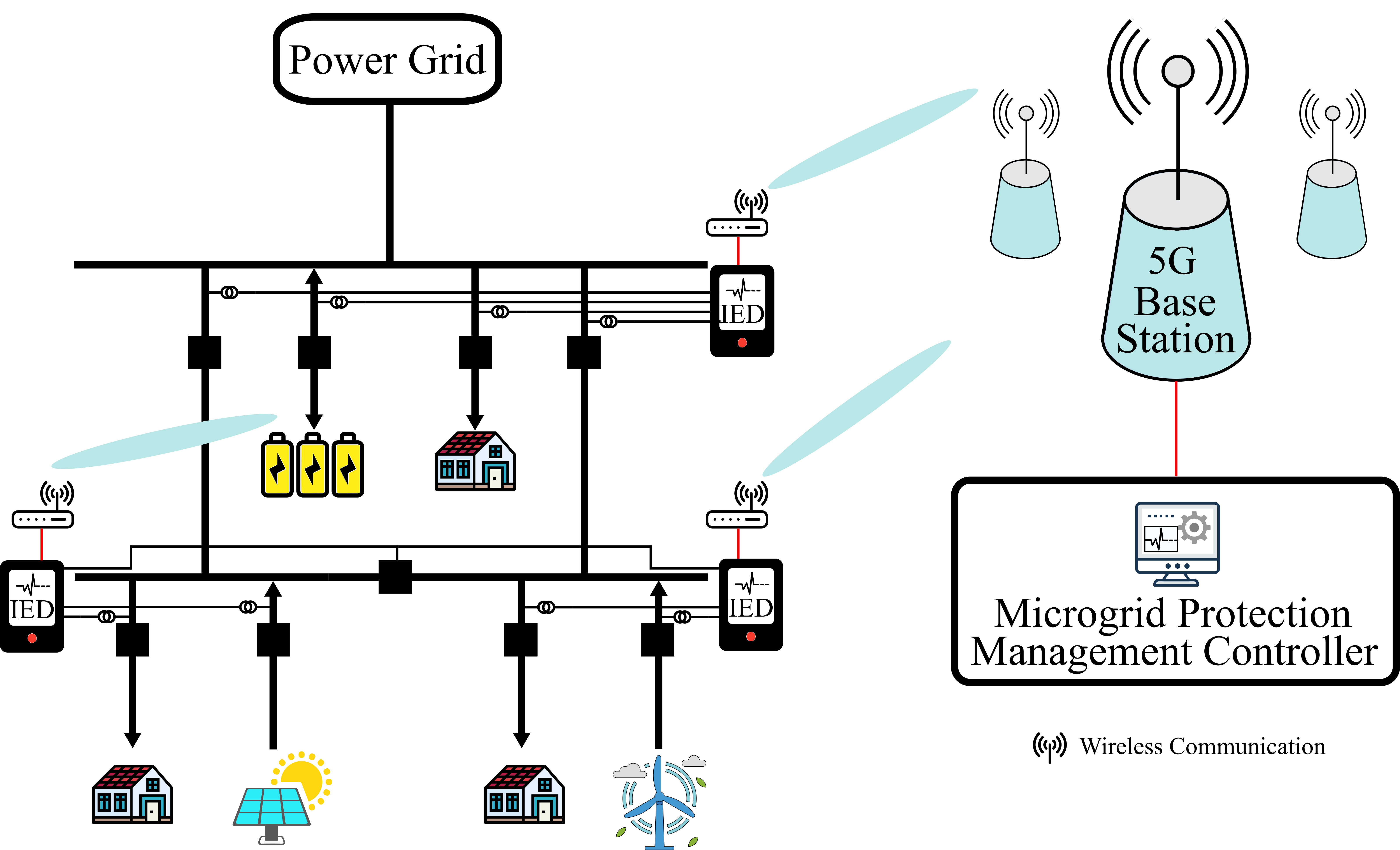}
    \caption{Wireless 5G communications deployment with interface diversity in an overcurrent adaptive protection scheme.}
    \label{fig:grid}
    \vspace{-2ex}
\end{figure}

Section \ref{sec:current_adaptive} showed that current microgrid sensoring and monitoring rely largely on wired communications, even though wireless systems can meet increasing quality of service requirements (as ongoing discussions on 5G suggest).
On a related note, the recent appearance of mobile 5G wireless communications, an evolution of 4G, as proposed by the latest realises of the 3rd Generation Partnership Project, has revealed highly promising for various vertical use cases, with reported efficient technical and economical solutions \cite{Popovski2018}. 
In the years to come, 5G networks shall include features targeted at improved performance for specific vertical use-cases (as the case of energy and automation verticals).
Advantages of 5G communication infrastructure includes cost savings (no wired physical connections are needed), network virtualization, improved response time, efficiency, flexibility, redundancy and its platform-approach, where a single interface is used to provide different types of connectivity \cite{Hovila2019}. 
Microgrid protection will eventually benefit from 5G technology developments, as it matures, since all network communications within the different elements from traditional protection or adaptive protection can be made using a centralized scheme as show in Fig. \ref{fig:grid}.

In particular, 5G is framed as having three cornerstones:
\begin{itemize}
    \item \textbf{Enhanced Mobile Broadband}: More data rate and connectivity than previous technology (4G);
    \item \textbf{Massive Machine Type Communication}: Larger number of devices connected than 4G and possibility of Machine-to-Machine communications;
    \item \textbf{\ac{urllc}}: 1 ms latency and 99,999\% reliability.
\end{itemize}

All the above features are relevant and will play a key role in substation control and grid automation. 
For instance, system operators can connect devices that are located in zones with difficult access.
5G would also allow for protection to become more distributed by installing IEDs at points closer to consumption, and DER generation without having to build new communication infrastructure. 
mMTC schemes could be used by IEDs to communicate  without having to rely on central servers for actuation purposes (e.g.,  reclosing schemes or informing the current state of a branch), as well as including one or multiples IEDs to the network, maintaining the same base stations (scalability). 
If one considers a large network deployment, as in a big city, massive connectivity between the elements is needed. 

However, URLLC is the most promising regime for adaptive protection in microgrids.
Previous work shows that message latency should be constrained by 2 cycles (i.e., 40 ms for a 50 Hz power system) \cite{Ustun2013}, while other indicates a stricter  requirement between 12 and 20 ms \cite{Yan2013}; both considering high reliability.
Current 4G systems can deliver an end-to-end latency of 20 ms, at best, which is a result of the constraint  from the frame structure. For example, tests in a 4G industrial private network achieved in the most favourable settings a delay of 26 ms (in comparison to a wired Ethernet scenario that achieved a delay of 3 ms)  \cite{polunin2019demonstrating}.
It is important to mention that, although 5G URLLC targets latencies as low as 1 ms, our particular application is less strict requiring 12 ms latency at the most stringent cases.

In terms of reliability, the performance of 4G is reliant upon several parameters, from the size of the message to the number of users.
In \cite{Nielsen2018}, the authors have proposed a quantitative relation between these key parameters based on field measurements.
The URLLC regime in 5G relates latency and reliability in a sense that the target reliability should be achieved within a very low latency constrain; originally, this constraint was 1 ms, but in the latter years it has been relaxed according to some more elaborate requirements for Industrial IoT (see, for example, \cite[Table 1]{5G-ACIA}).
Even these more relaxed versions, including the one we are using for the adaptive protection case, cannot be met by 4G.

In this case, the data-driven reliability guarantees based on a statistical learning framework seems a more suitable approach than the ``deterministic'' 1 ms potentially provided in URLLC regime \cite{Angje2019}. 
Depending on the application, ultra-reliability is critical but the low latency is more flexible; the adaptive protection exemplifies this.
Besides, recent results have proved that interface diversity where 5G combined with other wireless interfaces can provide ultra-reliability with bounded delay, which would satisfy the adaptive protection requirements \cite{Nielsen2018}.
%
%

%
%
%
The integration of these different quality of service can be done by NS, which is a concept that finds an efficient way for serving a determined application with 5G features on a common infrastructure \cite{Foukas2017,Kaloer2018}.
Various works in fields of communication for applications in Industry 4.0 show that NS using programmability and flexibility can be used to reduce complexity. This allows getting the best feature from a communication network, depending on the requirements from specific applications \cite{Popovski2018a}.
A slice can be considered an independent network, with corresponding advantages; in microgrid protection, it could be divided in many slices depending of the availability, latency or message type, as shown in \cite{Mendis2019}.
This concept makes communications even more flexible. As RES penetration increases in distribution systems, particularly in microgrids, the bidirectional fault current magnitudes become bigger, more sensors need to be installed, and therefore more signals need to be monitored. It then becomes a growing challenge for communication systems to deliver different messages from sensing devices to controllers and actuators. NS architectures may be able to efficiently deal with the complexity of handling such different and demanding requirements, which can range from high reliability and low latency to high data rates on the same industrial application.
%
%

All in all, new ways to incorporate wireless technology in substation automation and control need to be researched in the upcoming years, to accompany the rapid changes electrical distribution systems are already undergoing. The wired communication infrastructure will not be able to catch up, due to the lack of scalability and further prohibitive characteristics. A good approach would be multi-connectivity that combines wired and wireless, as those technologies have different failure patterns.
5G communications will open new frontiers in how these systems can be effectively integrated to perform tasks such as adaptive protection with very stringent requirements \cite{thrybom20155g}. 
In particular,  ultra-reliable communications with latency constraints required to perform adaptive protection should co-exist with other applications with multiple requirements, including massive connectivity of machine-type devices and more traditional broadband applications.
While current 5G solutions are not yet capable of reaching the demanded performance in protection applications, upcoming releases of 5G -- and even of 6G -- are expected to focus on specific vertical applications and application-specific requirements.
In this context, fast technological developments including potentially groundbreaking concepts, such as Semantic Filters \cite{Popovski2019} and Edge Intelligence, are expected to take place in the upcoming years\cite{park2019wireless}.
These developments should allow for tailored wireless communication solutions i.e. based on specific applications and their particular requirements, that co-exist and share the same  resources.
Usually, societal paradigm changes take place decades after key technologies (such as 5G) have been developed and rapid adoption is limited by conservative and progressive investment.
The adoption of wireless connectivity in energy sector has not yet become  mainstream. Some solutions like 4G and WiFi are deployed for some applications (mainly monitoring, metering and demand response), but not for adaptive protection, due to their performance limitations. Upgrading infrastructure to add 5G capabilities would bring additional capital costs considering incremental deployment in the existing grid infrastructure. On the other hand, it is expected that 5G brings down the operational costs related to communication network operations, due to its modularity and scalability  \cite{bag2017challenges}.
In 5G, the concept of local operator and private cellular networks indicates the tendency of third-party service providers, which is expected to decrease the operational costs related to the communication network, compared to more expensive deployment and maintenance  of wired networks  \cite{siriwardhana2019micro}.

5G has many potential advantages but also some challenges related to its effective implementation. These challenges are commonly associated with cyber-security. 
Careful examination of communication technologies has to be taken into consideration during a control and protection project design stage. The authors suggest this step to be essential for the economic viability of the project, since it can greatly reduce costs. This design should also include a robust system architecture to prevent or avoid possible cyber-attacks, given the vulnerability of wireless communication systems over wired communication systems. The reason for this is the wireless air propagation channel, where signals can be picked up from nearby locations without interfering in any hardware equipment.
%

It is worth restating that the proposed adaptive protection scheme can greatly reduce costs associated with communication network, bringing more flexibility in comparison to the traditional wired solutions. 
The benefits of using 5G would be also combined with already deployed solutions, leading to  gains from multi-connectivity, which is a popular way of attaining now that there are many wireless interfaces available \cite{Nielsen2018}. 
In summary, we argue that the proposed solution generally complies with the current deployments, which yields a smooth transition that will bring not only technical benefits but also economical ones.
\vspace{-1ex}
\section{Conclusion}
\label{sec:conclusions}

This paper presented key technical aspects related to the communication system that is needed to perform adaptive protection in micro-grids with high penetration of DERs. 
We particularly focused on different exiting solutions for adaptive protection systems, which are dominantly based on wired solutions.
We covered the traditional communication architectures (e.g., centralized or decentralized) and standards (e.g., GOOSE, SMV, RTPS among others).
We also discussed aspects related to cyber-security, including potential threats and types of attacks.
What is remarkable, though, is that current approaches mostly rely on wired networks despite the unquestionable performance gains of wireless technologies during the last decade.
In this sense, we argue that 5G in combination with other existing solutions (e.g., WiFi) can already achieve the required reliability of 99.999 \% with a bounded latency as low as 12 ms so that they should be seriously considered as a feasible enabler of adaptive protection applications.
In the near future, we expect that these solutions will take over many traditionally wired applications since wireless solutions tend to be cheaper, more flexible and easier to implement than wired ones to perform the same tasks, including mission-critical ones.
All in all, this review highlighted the state-of-the-art in the field indicating possible research directions that shall be taken to effectively deploy adaptive protection using wireless communications.

\section*{Acknowledgements}

This paper is partly supported by Academy of Finland via: (a) ee-IoT  n.319009, (b) FIREMAN consortium CHIST-ERA/n.326270, and (c) EnergyNet  Fellowship n.321265/n.328869.
The authors would like to thank the funding from DIGI-USER research platform

\vspace{-1ex}
\bibliographystyle{IEEEtran}
\bibliography{adaptive_protection_and_communications.bib}

\begin{IEEEbiography}[{\includegraphics[width=1in,height=1.25in,clip,keepaspectratio]{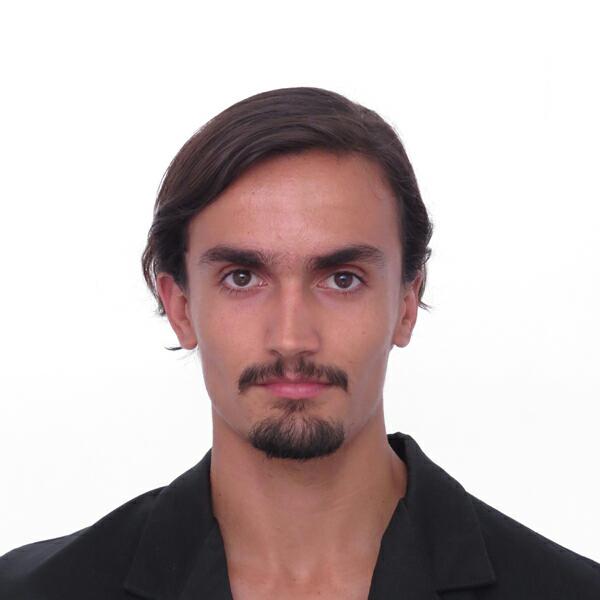}}]{Daniel Gutierrez-Rojas}
received the B.Sc. degree in Electrical Engineering from University of Antioquia, Colombia in 2016 and the M.Sc. degree in Protection of Power Systems University of São Paulo, Brazil, in 2017. From 2017 to 2019, he worked as security of operation and fault analyst for Colombia’s National electrical operator. He is currently working toward the Ph.D. degree at the School of Energy Systems at LUT University, Finland. His research interests include predictive maintenance, power systems, microgrids, mobile communication systems and electrical protection systems.
\end{IEEEbiography}

\begin{IEEEbiography}[{\includegraphics[width=1in,height=1.25in,clip,keepaspectratio]{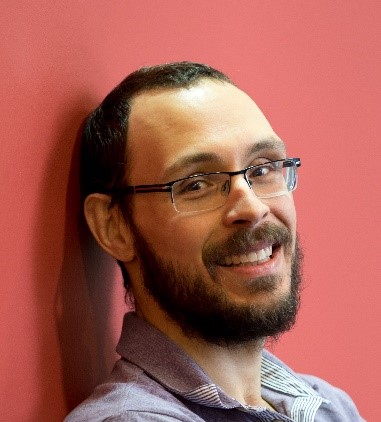}}]{Pedro H. J. Nardelli }
received the B.S. and M.Sc. degrees in electrical engineering from the State University of Campinas, Brazil, in 2006 and 2008, respectively. In 2013, he received his doctoral degree from University of Oulu, Finland, and State University of Campinas following a dual degree agreement. He is currently Assistant Professor (tenure track) in IoT in Energy Systems at LUT University, Finland, and holds a position of Academy of Finland Research Fellow with a project called Building the Energy Internet as a large-scale IoT-based cyber-physical system that manages the energy inventory of distribution grids as discretized packets via machine-type communications (EnergyNet). He leads the Cyber-Physical Systems Group at LUT and is Project Coordinator of the CHIST-ERA European consortium Framework for the Identification of Rare Events via Machine Learning and IoT Networks (FIREMAN). He is also Adjunct Professor at University of Oulu in the topic of “communications strategies and information processing in energy systems”. His research focuses on wireless communications particularly applied in industrial automation and energy systems. He received a best paper award of IEEE PES Innovative Smart Grid Technologies Latin America 2019 in the track “Big Data and Internet of Things”. He is also IEEE Senior Member. More information: https://sites.google.com/view/nardelli/
\end{IEEEbiography}

\begin{IEEEbiography}[{\includegraphics[width=1in,height=1.25in,clip,keepaspectratio]{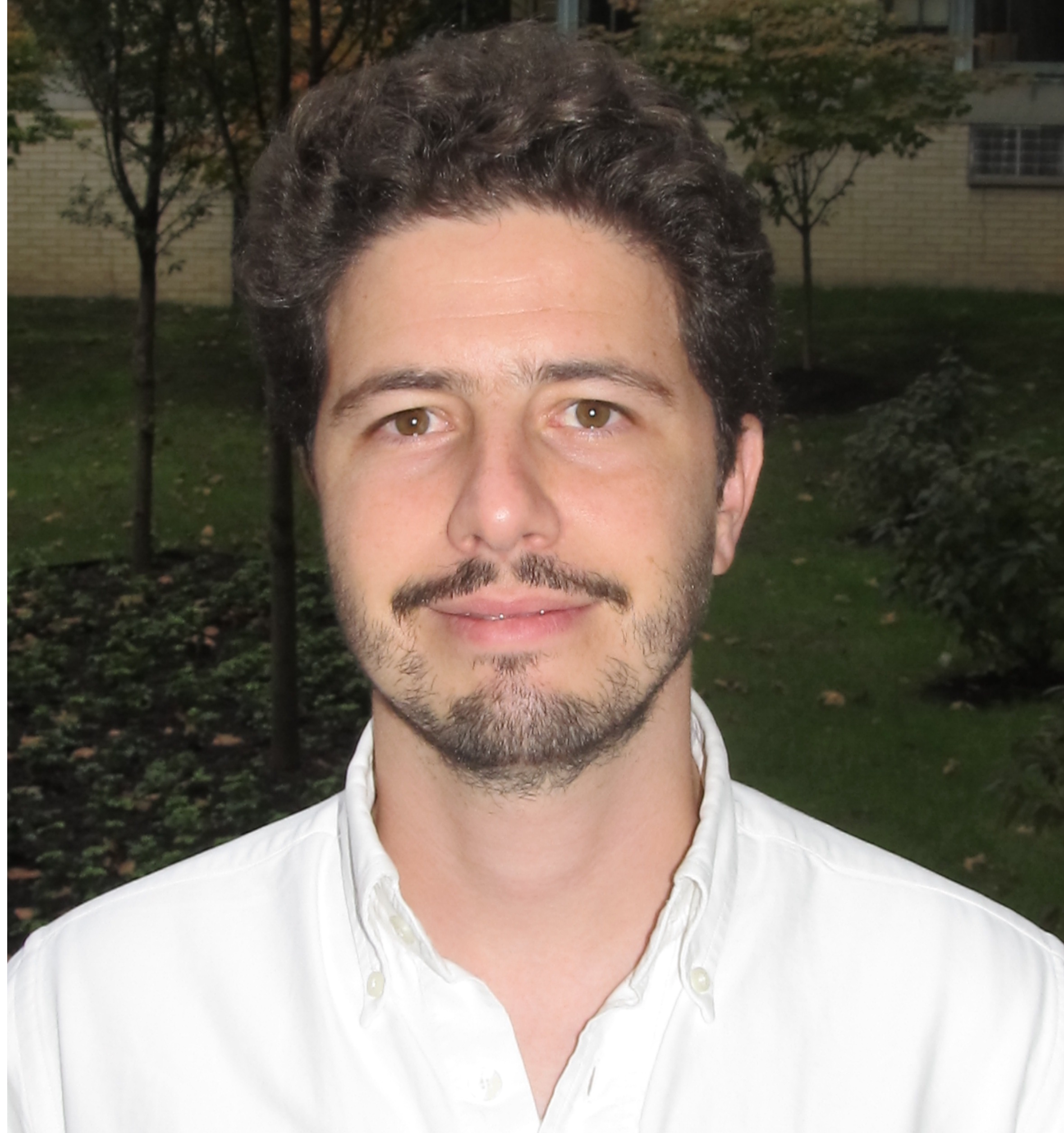}}]{Goncalo Mendes}
is a Postdoctoral researcher at the LUT School of Energy Systems, in Finland. His main research interests lie on the intersection of local energy systems with policy and business aspects. He has recently received a Fulbright scholarship to investigate regulatory challenges faced by multi-stakeholder microgrid projects and to derive novel enabling policies from these lessons. Dr. Mendes received his Ph.D. in Sustainable Energy Systems (MIT Portugal Program) from the University of Lisbon in early 2017. He is a representative for Europe at the International Steering Committee of the annual Microgrid Symposiums and a member of CIRED’s Working Group on Microgrid Business Models and Regulatory Issues.
\end{IEEEbiography}

\begin{IEEEbiography}[{\includegraphics[width=1in,height=1.25in,clip,keepaspectratio]{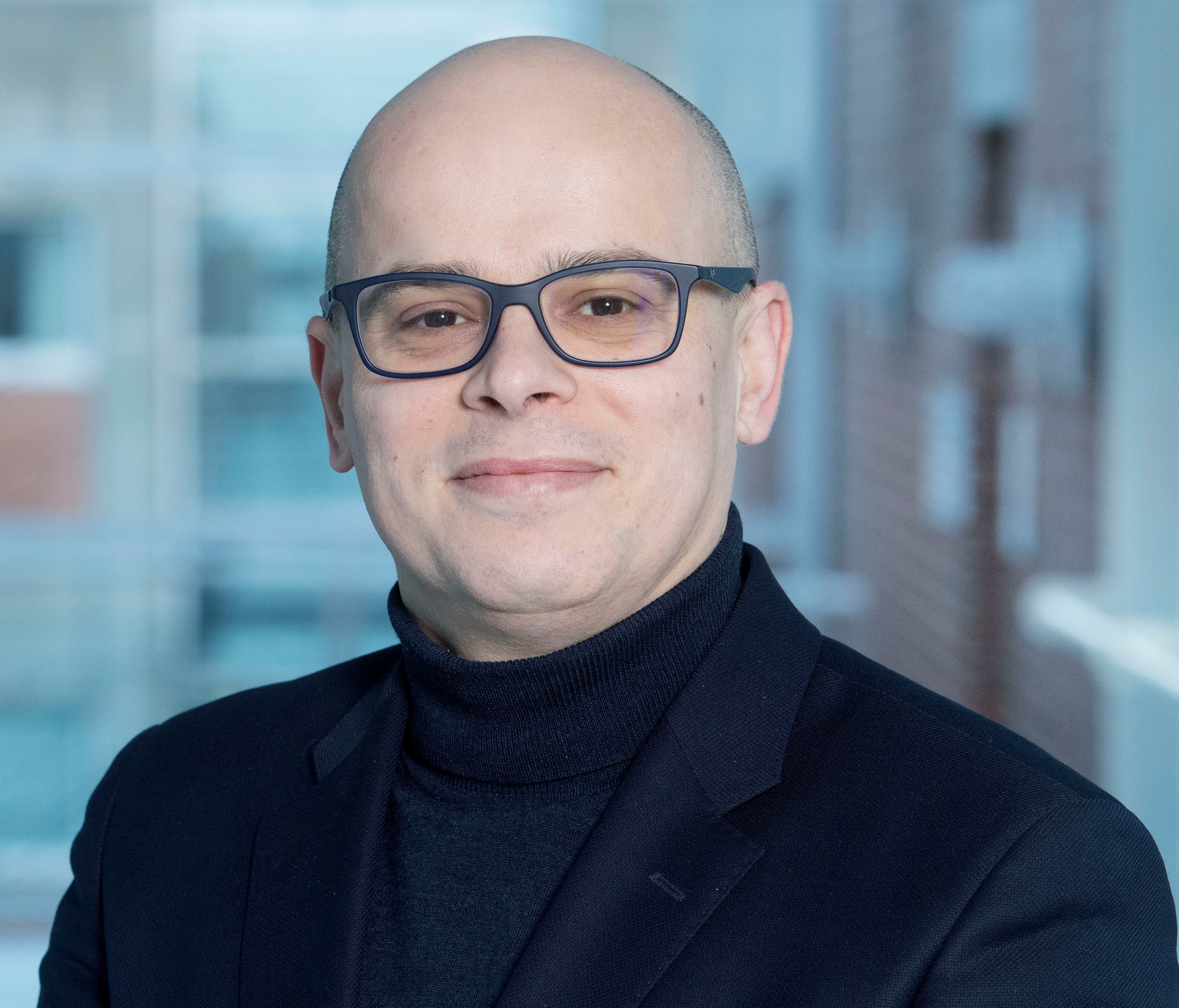}}]{Petar Popovski}
(S'97--A'98--M'04--SM'10--F'16) is a Professor at Aalborg University, where he is heading the section on Connectivity. He received his Dipl. Ing and M. Sc. degrees in communication engineering from the University of Sts. Cyril and Methodius in Skopje and the Ph.D. degree from Aalborg University in 2005. He is a Fellow of IEEE, has over 300 publications in journals, conference proceedings, and edited books and he was featured in the list of Highly Cited Researchers 2018, compiled by Web of Science. He holds over 30 patents and patent applications. He received an ERC Consolidator Grant (2015), the Danish Elite Researcher award (2016), IEEE Fred W. Ellersick prize (2016), IEEE Stephen O. Rice prize (2018) and the Technical Achievement Award from the IEEE Technical Committee on Smart Grid Communications. He is currently a Steering Committee Member of IEEE SmartGridComm and IEEE Transactions on Green Communications and Networking. He previously served as a Steering Committee Member of the IEEE INTERNET OF THINGS JOURNAL. He is currently an Area Editor of the IEEE TRANSACTIONS ON WIRELESS COMMUNICATIONS. Prof. Popovski was the General Chair for IEEE SmartGridComm 2018 and IEEE Communication Theory Workshop 2019. From 2019, he is also a Member-at-Large of the Board of Governors of the IEEE Communications Society. His research interests are in the area of wireless communication, communication theory and Internet of Things. In 2020 he published the book "Wireless Connectivity: An Intuitive and Fundamental Guide".
\end{IEEEbiography}


%

\end{document}